# "The Roller Conduction Effect" from the A-share Data Evidence

Wenbo Lyu

*Saxo Fintech Business School, University of Sanya, Sanya, 572000, China*
*Tel/Fax: +86 18617571856; E-mail:wenbolv@sanyau.edu.cn*

**Abstract:** In the post-epidemic era, consumption recovery has obvious time and space transmission laws, and there are different valuation criteria for consumption segments. Using the A-share data of the consumption recovery stage from January to April 2022, this paper quantitatively compares the rotation effect between different consumption sectors when the valuation returns to the reasonable range. According to the new classification of "sensory-based consumption", it interprets the internal logic of digital consumption as A consumption upgrade tool and a higher valuation target, and expounds the "the roller conduction effect". The law of consumption recovery and valuation return period is explained from the perspective of time and space conduction. The study found that in the early stage of consumption recovery, the recovery of consumer confidence was slow. In this period, A-shares were mainly dominated by the stock capital game, and there was an obvious plate rotation law in the game. Being familiar with this law has strong significance, which not only helps policy makers to adjust the direction of policy guidance, but also helps financial investors to make better investment strategies. The disadvantage of this paper is that it has not yet studied the roller conduction effect of the global financial market, and more rigorous mathematical models are still needed to support the definition of stock funds, which is also the main direction of the author's future research.

## 1. Introduce

In the context of China's economic performance in 2022 and Chinese government policies in 2023, the internal logic of digital consumption as a tool for consumption upgrading and a higher valuation goal is explained, leading to an investment strategy with a roll-on effect. Wang Qing, Wang Zhongli et al. indicated that the impact of the epidemic on stock prices had a certain lag effect, but the lag effect lasted only 1 day. Investor fear and anxiety caused by the epidemic could spur investors to sell stocks, leading to volatility in the stock market. In 2023, the government will continue to improve the market environment, improve the governance structure of listed companies, introduce more quality listed companies, and promote the implementation of more investor protection policies. In addition, the government will also promote the deepening reform of the capital market, improve and perfect the regulatory system of the capital market, improve the market environment, enhance investor confidence, and promote the healthy development of the capital market.

Based on the law of consumption recovery transmission in the post-epidemic era, the definition of sensory consumption is given, and the valuation rules of consumer companies are refined. From the perspective of investors, the "digital consumption" is given a higher valuation, and the "barrel effect" is proposed in the recovery transmission mechanism. And based on China's A-share panel data from January 1, 2023 to the present, the empirical example of the barrel effect is given.

**2. Concept definition**
The utility of people's consumption of products and services can be classified according to somatosensory, and all the utility will be implemented to the satisfaction of one or several senses. The marginal utility of the five senses of the audiovisual touch is different. The valuation rules of different sensory consumption sectors are different, and stock price anomalies are more likely to occur in the sectors with high valuation.

Based on the experience of valuation weights and stock price anomalies:
Visual sensation > auditory sensation > olfactory sensation > tactile sensation > taste sensation;
Comprehensive service sense to see the specific synthesis of senses to superimpose weight ranking, such as mobile phone digital class, superimposed visual sense, hearing sense, body (touch) sense of multiple senses, so it is easier to high valuation, but also easier to produce stock price visions. Through the observation of consumption in developed countries, we find that the influence of digital consumption is very huge in the entire consumer sector.

The following table is a comparison of the time series data of the overall consumption plate and the electronic consumption (digital consumption in this paper) plate in the A-share market.

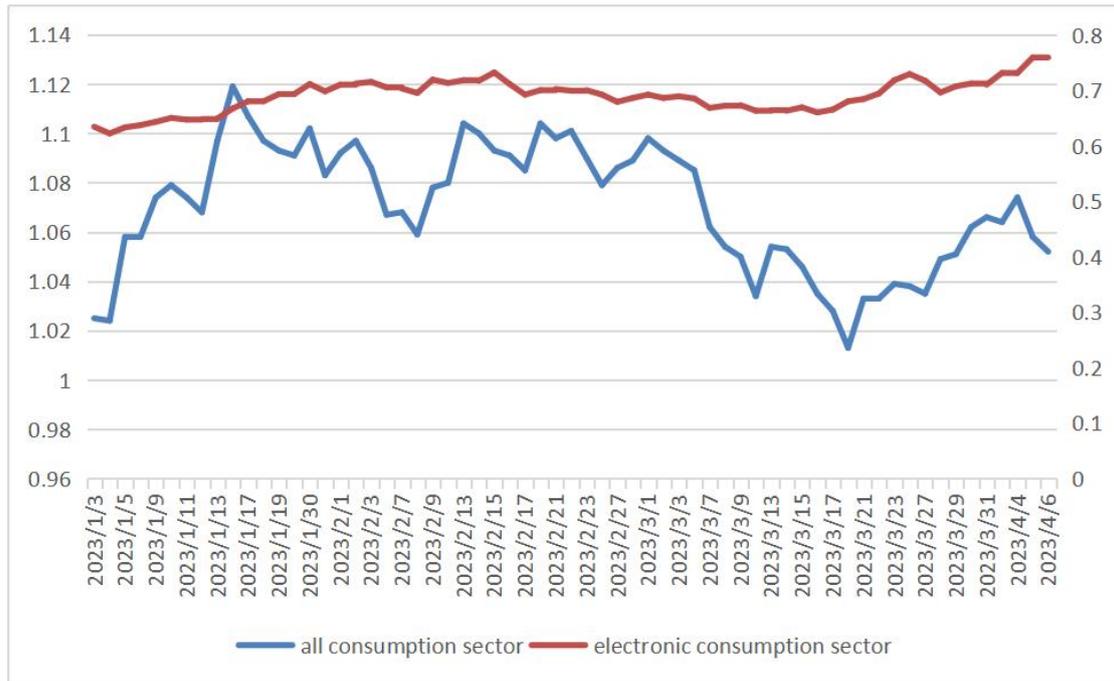

**Figure 1.** Trends of overall consumption sector and electronic consumption sector

The stock market is a barometer of the real economy and is usually thought to be half a year ahead. From January 3, 2023 to now, the consumption sector rose by 2.63%, the trend appeared first up and then down, suggesting that the recovery of consumption in the real economy is weak, the demand is less than expected; Consumer electronics sector rose 19.56 percent, suggesting that in the weak consumer demand, electronics supply and demand are booming, faster recovery, stronger trend. The supporting evidence is given to the viewpoint of this paper.

**3.The roller conduction effect**
The trend of financial asset prices can hardly be predicted perfectly, but according to historical experience, price fluctuations, especially the process of reasonable valuation regression, have certain rules.

In the previous academic work, the author proposed the "roller conduction effect". Assuming that the sum of funds traded on the stock market and on the sidelines is relatively fixed for a period of time, the overall volume of these funds is not enough to instantaneously make all reasonably valued stocks rise simultaneously, and maintain the same rising rate; Secondly, the decision-makers of funds represent long-term style funds, short-term style funds, emotional style funds and regulators will funds, and the decisions of different styles are impossible to be consistent. So we can say that even if a plate or several plates in the market (including horizontal and vertical) have a far-end consistent expectation, there will be a near-end asynchronous rotation effect, superimpose hot money and retail emotional factors, will enhance the amplitude of this rotation, according to the time has shown obvious between the plate, the strong and weak rules.

It is worth noting that the barrel conduction effect is suitable for the game prediction of stock funds in A period of time, if the limit of this period is broken, or there is a huge amount of funds into or out, such as a huge amount of northbound funds into the A-share market, it will break through the "barrel wall" rule, resulting in a great change in the roller conduction effect.

Investors, in an incomplete market, should not chase the momentum of water in the barrel (the frequent trading state of chasing up and down), the wisest thing is to patiently hold certain assets and wait for the roller effect to realize profits.

**4. Model construction**
4.1 Coordination degree model
The energy relationship and coordination algorithm between electrical circuits are extended to the relationship comparison between different economic systems. Currently, the coupling coordination degree model has been widely used in the academic circle to quantify the degree of correlation and mutual influence between multiple systems. In order to better study the coupling coordination between the four stock index plates, four coupling degree models need to be built based on the actual situation. The formula is as follows:

$$C = \left\{ \frac{U_1 \times U_2 \times U_3 \times U_4}{\left[\frac{U_1 + U_2 + U_3 + U_4}{4}\right]^4} \right\}^{\frac{1}{4}} = \frac{4\sqrt[4]{U_1 \times U_2 \times U_3 \times U_4}}{U_1 + U_2 + U_3 + U_4}$$

Among them, c in the formula is the coupling degree, and the greater the value of c, the higher the coupling degree between the systems. When the value of C is 1, it indicates that the subsystems within the system are in a high degree of orderly development. Arc refers to the comprehensive development level index among innovation operation subsystem, innovation research and development subsystem, innovation support subsystem and innovation environment subsystem respectively.

(2) Comprehensive development level index
According to the coupling degree model, the coupling degree of the system is related to the comprehensive development level index of each subsystem. The level index U can effectively reflect the development level and relative development degree of each subsystem, and reflect the contribution degree of indicators in the system to the overall function of the system. Before the calculation, the default weight ratio of each index is the same.

(3) Innovation ecosystem coupling coordination degree model
The coupling degree model can reflect the degree of coupling between systems, but when the development level of multiple systems is high or low at the same time

When, it will also show the phenomenon of high overall coupling degree, and there are certain drawbacks. Therefore, a more scientific and rigorous coupling coordination degree model is introduced to further measure the degree of collaborative innovation between systems.

$$D = (C \times T)^{\frac{1}{2}}$$
$$\begin{cases} T = \alpha U_1 + \beta U_2 + \chi U_3 + \delta U_4 \\ \alpha = \beta = \chi = \delta = 0.25 \end{cases}$$

D represents the coupling coordination degree of innovation ecosystem, T represents the comprehensive influence degree of the comprehensive development level index of each subsystem on the overall coordination degree, α、β、χ、δ are undetermined coefficients, and the four subsystems of innovation ecology are set as equally important, so they are all equal to 0.25, and the sum is 1.

| the coupling degree and coordination degree | Values | stage |
|---|---|---|
| **the coupling degree** | 0<C≤0.3 | The initial coupling stage |
|  | 0.3<C≤0.5 | Low-level coupling stage |
|  | 0.5<C≤0.8 | Intermediate coupling stage |
|  | 0.8<C≤1 | Advanced coupling stage |
| **coordination degree** | 0<D≤0.3 | Initial coordination stage |
|  | 0.3<D≤0.5 | Low-level coordination stage |
|  | 0.5<D≤0.8 | Intermediate coordination stage |
|  | 0.8<D≤1 | Advanced coordination stage |

**Table 1.** Based on the definition of the range of coordination degree, the coupling degree and coordination degree are divided into four levels respectively

**5. Empirical analysis**
5.1 Examples of the roller conduction effect of China's A-shares: the correlation between SSE Index, SZI, GEI and STAR 50

Overall capital volume assessment: Using the change data curve of the market value of each section, it can be seen that from mid-January to April 2023, the total market value of the four major indexes has fluctuated between 8.6 trillion and 8.9 trillion, with obvious characteristics of stock capital status, and the total amount of capital has not experienced significant outflow or inflow.

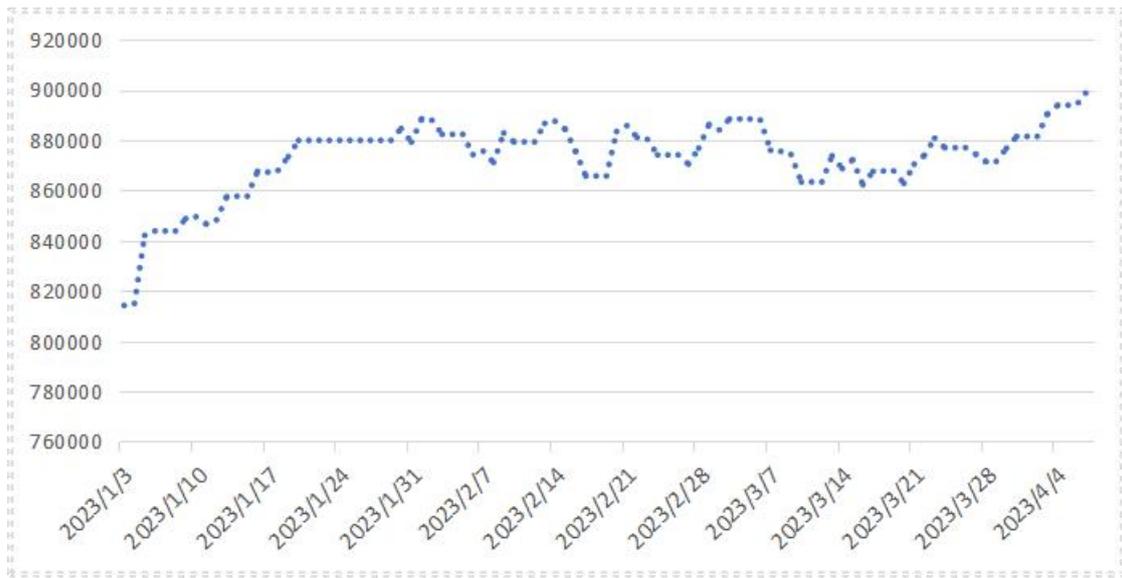

**Figure 2.** Aggregate of four indexes GMV(RMB)

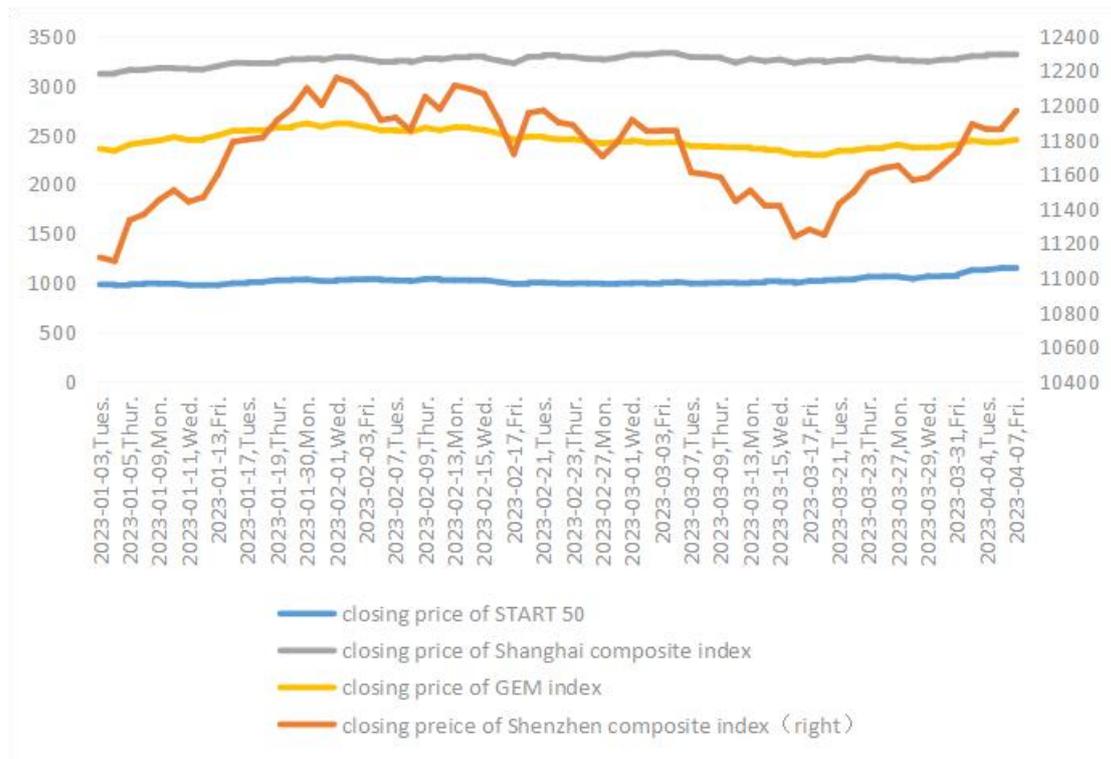

**Figure 3.** Trends of Shanghai Composite Index, Shenzhen Composite Index, GEM index, STAR 50

As can be seen from Figure 2, from January 3 to February 1, the four indexes all rose, showing synergy; From February 1 to March 15, the Shanghai Composite Index performed stably, while the other three indexes all fell significantly, showing the obvious game of stock funds. Investors transferred funds into the constituent plates of the Shanghai Composite Index, and the roller conduction effect showed initial signs. After March 15, the four indexes all rose significantly, among which the STAR 50 index rose the most sharply. The Science and Technology Innovation Board and the

Growth Enterprise Board have a certain degree of complementarity. The Science and Technology Innovation Board is mainly composed of emerging high-tech companies, while the Growth Enterprise Board is mainly composed of large mature high-tech companies. Assuming that the intervention of incremental capital is not taken into account, the rising amplitude of the two will gradually converge, and the capital will gradually flow from the science and innovation sector to the entrepreneurial sector and other index sectors.

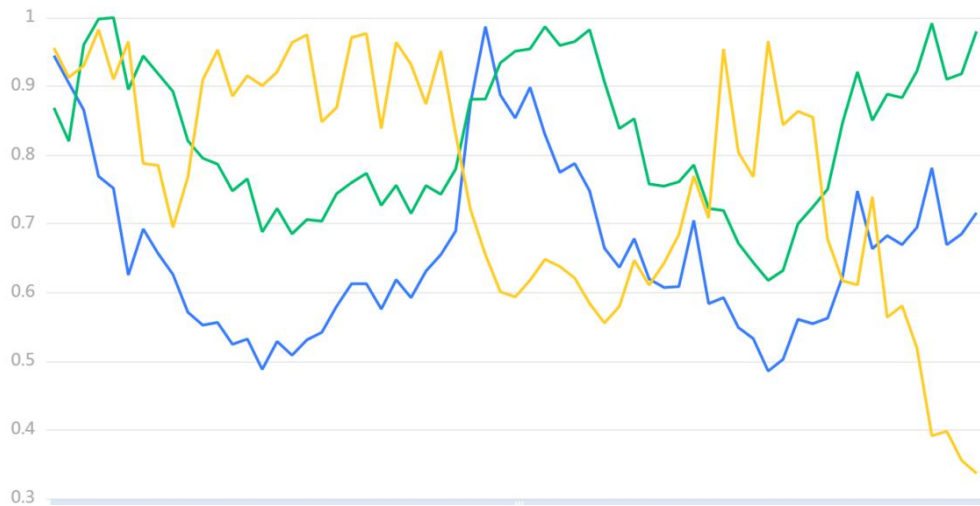

**Figure 4**.Negative correlation between STAR 50 (yellow) and Shenzhen Component index (green) and GEM index (blue)

As can be seen from Figure 3, with the trend correlation analysis (based on grey correlation analysis) and Shanghai Composite Index as the parent sequence, STAR 50 index has an obvious reverse change trend, which is in sharp contrast with Shenzhen component index and GEM index, indicating that stock funds have been transferred from GEM index to STAR 50.

Then, the coupled coordination degree of the four indexes is studied. The Shanghai Composite Index is still used as the benchmark, and the D-value trend of the coupled coordination degree of the three indexes is as follows.

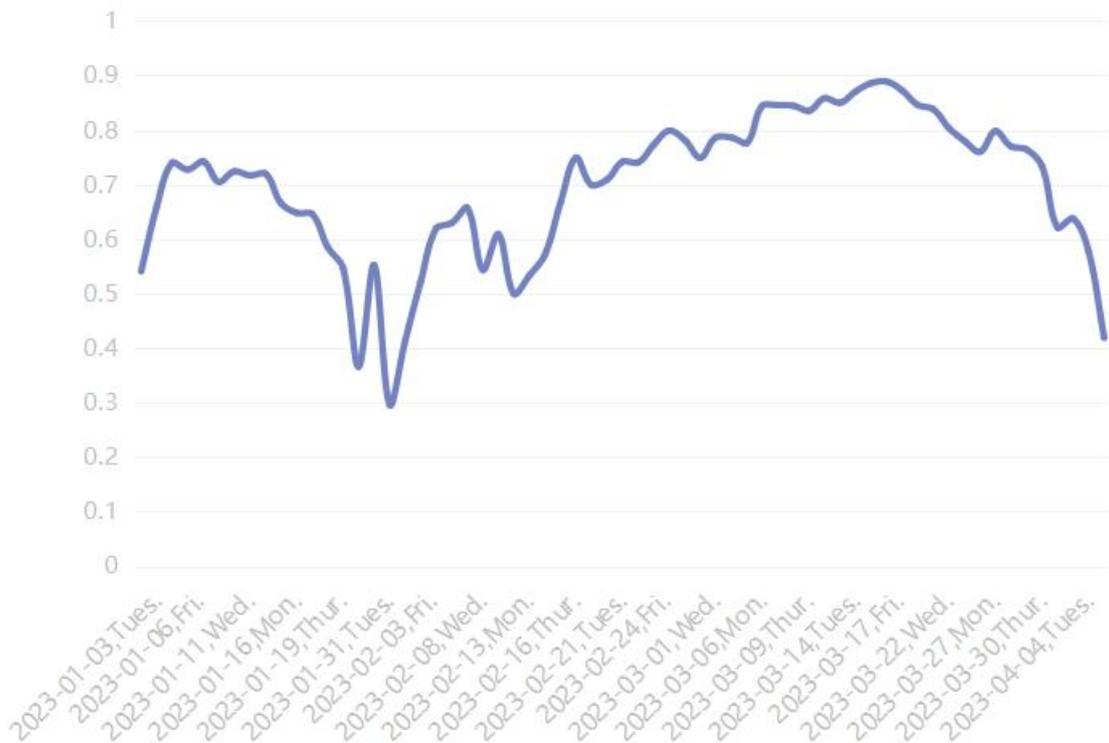

**Figure 5.** Coupling coordination degree D value

As of the beginning of April, the coupling coordination degree of STAR 50 and other plates is in a mismatched low coupling state, with a score of only 0.41 points, which also shows that the rolling barrel effect is obvious during the valuation recovery period, with a large rolling range of funds and a long cycle, and the coordinated change trend of each plate is weak.

## 6. Conclusions and findings

Through the data analysis of the closing price and total market value of the four major A-share indexes, this paper proves the phenomenon of "the roller conduction effect effect" in the stock market price transmission.

In the game of stock funds, the roller conduction effect is more significant. Generally speaking, when the stock price has a large pullback due to cyclical fluctuations, the stock index can finally rebound.

## 7.Appendix data and results：（Table 2）

| Date | SSE Index GMV (RMB 100 million) | STAR 50 GMV (RMB 100 million) | GEI GMV (RMB 100 million) | SZI GMV (RMB 100 million) | Aggregate of four indexes GMV (RMB 100 million) |
| --- | --- | --- | --- | --- | --- |

| 日期 | | | | | |
|---|---|---|---|---|---|
| 2023/1/3 | 505534.1655 | 27654.2025 | 58531.8780 | 222484.4611 | 814204.7071 |
| 2023/1/4 | 507178.5504 | 27460.6738 | 58095.4355 | 222255.3862 | 814990.0459 |
| 2023/1/5 | 527759.7458 | 27711.7211 | 59681.8849 | 226938.1169 | 842091.4687 |
| 2023/1/6 | 527989.9217 | 27958.1995 | 60185.5268 | 227722.4531 | 843856.1011 |
| 2023/1/7 | 527990.4556 | 27958.1995 | 60185.5873 | 227722.5137 | 843856.7561 |
| 2023/1/8 | 527990.4556 | 27958.1995 | 60185.5873 | 227722.5137 | 843856.7561 |
| 2023/1/9 | 530856.1915 | 27917.1028 | 60634.3491 | 229361.4531 | 848769.0965 |
| 2023/1/10 | 529511.2570 | 28002.4359 | 61533.8175 | 230442.7724 | 849490.2828 |
| 2023/1/11 | 529096.6697 | 27514.3085 | 60844.5626 | 229155.7859 | 846611.3267 |
| 2023/1/12 | 530062.6162 | 27464.8568 | 61125.4770 | 229593.7811 | 848246.7311 |
| 2023/1/13 | 535348.8555 | 27526.9618 | 62081.0999 | 232531.2020 | 857488.1192 |
| 2023/1/14 | 535538.9456 | 27526.9618 | 62081.0999 | 232531.2020 | 857678.2093 |
| 2023/1/15 | 535538.9456 | 27526.9618 | 62081.0999 | 232531.2020 | 857678.2093 |
| 2023/1/16 | 540230.0316 | 28075.5461 | 63211.9981 | 236058.5608 | 867576.1366 |
| 2023/1/17 | 539548.1742 | 28306.8792 | 63339.0055 | 236075.4340 | 867269.4929 |
| 2023/1/18 | 540087.0922 | 28329.0064 | 63296.5045 | 236258.1357 | 867970.7388 |
| 2023/1/19 | 542634.3896 | 28802.6250 | 63904.2903 | 238115.1914 | 873456.4963 |
| 2023/1/20 | 547362.1832 | 28923.3514 | 64173.5386 | 239441.1324 | 879900.2056 |
| 2023/1/21 | 547407.2668 | 28923.3514 | 64173.5386 | 239441.1324 | 879945.2892 |
| 2023/1/22 | 547407.2668 | 28923.3514 | 64173.5386 | 239441.1324 | 879945.2892 |
| 2023/1/23 | 547407.2668 | 28923.3514 | 64173.5386 | 239441.1324 | 879945.2892 |
| 2023/1/24 | 547407.2668 | 28923.3514 | 64173.5386 | 239441.1324 | 879945.2892 |
| 2023/1/25 | 547407.2668 | 28923.3514 | 64173.5386 | 239441.1324 | 879945.2892 |
| 2023/1/26 | 547407.2668 | 28923.3514 | 64173.5386 | 239441.1324 | 879945.2892 |
| 2023/1/27 | 547407.2668 | 28923.3514 | 64173.5386 | 239441.1324 | 879945.2892 |
| 2023/1/28 | 547407.2668 | 28923.3514 | 64173.5386 | 239441.1324 | 879945.2892 |
| 2023/1/29 | 547407.2668 | 28923.3514 | 64173.5386 | 239441.1324 | 879945.2892 |
| 2023/1/30 | 549067.6655 | 28916.5365 | 64880.2731 | 241756.6716 | 884621.1467 |
| 2023/1/31 | 546848.5938 | 28475.1451 | 63994.0950 | 239920.5883 | 879238.4222 |
| 2023/2/1 | 551731.2942 | 28900.5088 | 64750.5598 | 243093.5091 | 888475.8719 |
| 2023/2/2 | 551692.5523 | 29038.4581 | 64545.3856 | 242639.3470 | 887915.743 |
| 2023/2/3 | 548332.2655 | 29113.5905 | 63962.1251 | 240909.3928 | 882317.3739 |
| 2023/2/4 | 548333.5782 | 29113.5905 | 63962.1251 | 240909.3928 | 882318.6866 |
| 2023/2/5 | 548333.5782 | 29113.5905 | 63962.1251 | 240909.3928 | 882318.6866 |
| 2023/2/6 | 544319.8107 | 28888.6082 | 63067.6458 | 237963.9074 | 874239.9721 |
| 2023/2/7 | 545584.9276 | 28744.9458 | 62966.6320 | 238294.8477 | 875591.3531 |
| 2023/2/8 | 542975.5228 | 28430.7270 | 62692.7744 | 237029.9015 | 871128.9257 |
| 2023/2/9 | 549304.5951 | 29178.7882 | 63743.1066 | 240629.8503 | 882856.3402 |
| 2023/2/10 | 547850.2857 | 28907.0768 | 63224.4264 | 239317.7142 | 879299.5031 |
| 2023/2/11 | 547862.8847 | 28907.0768 | 63224.4264 | 239317.6566 | 879312.0445 |
| 2023/2/12 | 547862.8847 | 28907.0768 | 63224.4264 | 239317.6566 | 879312.0445 |
| 2023/2/13 | 551379.9523 | 28935.7801 | 64079.2242 | 242265.8246 | 886660.7812 |
| 2023/2/14 | 552774.9415 | 28899.2973 | 63872.7550 | 241991.7806 | 887538.7744 |
| 2023/2/15 | 550697.7607 | 28986.8070 | 63444.7939 | 241154.2358 | 884283.5974 |

| Date | | | | | |
|---|---|---|---|---|---|
| 2023/2/16 | 546443.4951 | 28417.6981 | 62590.5625 | 238068.1308 | 875519.8865 |
| 2023/2/17 | 542158.3425 | 27874.1899 | 61117.9777 | 234560.8168 | 865711.3269 |
| 2023/2/18 | 542167.8120 | 27874.3067 | 61117.9777 | 234560.8168 | 865720.9132 |
| 2023/2/19 | 542167.8120 | 27874.3067 | 61117.9777 | 234560.8168 | 865720.9132 |
| 2023/2/20 | 554055.7776 | 28194.7864 | 61913.5983 | 239280.1588 | 883444.3211 |
| 2023/2/21 | 556532.1567 | 28109.5119 | 61556.0715 | 239559.7594 | 885757.4995 |
| 2023/2/22 | 553698.8671 | 27952.0385 | 61169.3187 | 238270.6028 | 881090.8271 |
| 2023/2/23 | 553090.1817 | 28027.8329 | 61240.0042 | 237950.0666 | 880308.0854 |
| 2023/2/24 | 549676.8429 | 27988.2011 | 60537.0967 | 235911.4983 | 874113.639 |
| 2023/2/25 | 549676.8725 | 27988.2011 | 60538.4021 | 235912.7832 | 874116.2589 |
| 2023/2/26 | 549676.8725 | 27988.2011 | 60538.4021 | 235912.7832 | 874116.2589 |
| 2023/2/27 | 548266.0880 | 27860.5341 | 60045.0598 | 234305.5918 | 870477.2737 |
| 2023/2/28 | 552400.6764 | 27986.9847 | 60532.4052 | 235896.2262 | 876816.2925 |
| 2023/3/1 | 558819.5369 | 28078.7431 | 60929.1676 | 238416.2977 | 886243.7453 |
| 2023/3/2 | 558872.9335 | 27844.1993 | 60263.4996 | 237120.8556 | 884101.488 |
| 2023/3/3 | 562728.0570 | 28117.8476 | 60333.4158 | 237229.5588 | 888408.8792 |
| 2023/3/4 | 562728.7896 | 28117.8476 | 60333.4158 | 237229.5588 | 888409.6118 |
| 2023/3/5 | 562728.7896 | 28117.8476 | 60333.4158 | 237229.5588 | 888409.6118 |
| 2023/3/6 | 561996.8939 | 28430.3054 | 60546.1823 | 236996.0303 | 887969.4119 |
| 2023/3/7 | 556248.2684 | 27867.2581 | 59361.8442 | 232507.4331 | 875984.8038 |
| 2023/3/8 | 556176.3190 | 28000.3343 | 59212.4367 | 232155.5869 | 875544.6769 |
| 2023/3/9 | 555400.1035 | 28102.8761 | 59059.6536 | 231711.3172 | 874273.9504 |
| 2023/3/10 | 547525.0061 | 27963.9771 | 59025.3633 | 228844.9081 | 863359.2546 |
| 2023/3/11 | 547557.9306 | 27963.9771 | 59025.3633 | 228845.1164 | 863392.3874 |
| 2023/3/12 | 547557.9306 | 27963.9771 | 59025.3633 | 228845.1164 | 863392.3874 |
| 2023/3/13 | 555917.7978 | 29120.9346 | 58698.0251 | 230022.9782 | 873759.7357 |
| 2023/3/14 | 552510.3167 | 29632.2548 | 58295.7680 | 228253.6562 | 868691.9957 |
| 2023/3/15 | 556150.7174 | 29499.4495 | 58143.5074 | 228368.6029 | 872162.2772 |
| 2023/3/16 | 550778.8970 | 29032.7359 | 57398.0498 | 225052.8410 | 862262.5237 |
| 2023/3/17 | 555221.1613 | 29689.8138 | 57197.5817 | 225463.9554 | 867572.5122 |
| 2023/3/18 | 555379.1268 | 29689.8138 | 57197.6126 | 225463.3901 | 867729.9433 |
| 2023/3/19 | 555379.1268 | 29689.8138 | 57197.6126 | 225463.3901 | 867729.9433 |
| 2023/3/20 | 551076.9757 | 29864.6686 | 57138.1325 | 224696.5454 | 862776.3222 |
| 2023/3/21 | 553837.3904 | 30132.9332 | 58301.7212 | 228162.6603 | 870434.7051 |
| 2023/3/22 | 555904.5556 | 30229.5750 | 58455.2244 | 229409.1271 | 873998.4821 |
| 2023/3/23 | 559694.5786 | 30734.2492 | 58955.1250 | 231428.1439 | 880812.0967 |
| 2023/3/24 | 555373.0524 | 30690.6168 | 59096.4869 | 231812.8341 | 876972.9902 |
| 2023/3/25 | 555376.5946 | 30690.6168 | 59096.4869 | 231812.8341 | 876976.5324 |
| 2023/3/26 | 555376.5946 | 30690.6168 | 59096.4869 | 231812.8341 | 876976.5324 |
| 2023/3/27 | 552275.1039 | 30548.5778 | 59724.7845 | 231878.6480 | 874427.1142 |
| 2023/3/28 | 551590.8622 | 30082.5775 | 59065.6142 | 230698.4491 | 871437.503 |
| 2023/3/29 | 550707.2594 | 30742.3705 | 59104.4677 | 230997.4573 | 871551.5549 |
| 2023/3/30 | 554001.2640 | 30815.3631 | 59434.4513 | 232533.8104 | 876784.8888 |
| 2023/3/31 | 556659.4282 | 31168.9071 | 59806.6558 | 233846.1386 | 881481.1297 |

| 2023/4/1 | 556660.4798 | 31169.6685 | 59806.6558 | 233846.1054 | 881482.9095 |
| 2023/4/2 | 556660.3712 | 31169.6685 | 59806.6558 | 233846.1054 | 881482.8009 |
| 2023/4/3 | 561172.1159 | 32453.0414 | 60556.1012 | 236355.7747 | 890537.0332 |
| 2023/4/4 | 565356.4233 | 32558.4258 | 60071.9418 | 235916.8701 | 893903.661 |
| 2023/4/5 | 565369.6658 | 32558.4258 | 60071.9419 | 235916.8817 | 893916.9152 |
| 2023/4/6 | 565900.0463 | 33018.5746 | 60135.9509 | 235876.4018 | 894930.9736 |
| 2023/4/7 | 568474.9299 | 33395.9586 | 60667.2947 | 237774.6505 | 900312.8337 |